\documentstyle[preprint,aps,prb,psfig]{revtex}

\tighten

\include{epsfig}

\begin{document}

\draft

\preprint{CD-06}

\title{Hysteresis loop areas in kinetic Ising models:
       Effects of the switching mechanism}

\date{\today}

\author{S.W. Sides,$^{\ast \dagger \ddagger}$
        P.A. Rikvold,$^{\ast \dagger \ddagger}$
        and M.A. Novotny $^{\dagger \star}$}

\address{$^{\ast}$Center for Materials Research
and Technology and Department of Physics, and \\
$^{\dagger}$Supercomputer Computations Research Institute, \\
Florida State University, Tallahassee, Florida 32306-4130 \\
$^{\ddagger}$Colorado Center for Chaos and Complexity,
University of Colorado, Boulder, Colorado 80309-0216 \\
$^{\star}$Department of Electrical Engineering,
{FAMU-FSU College of Engineering, \\
Pottsdamer Street, Tallahassee, Florida 32310-6046}}

\maketitle

\begin{abstract}
Experiments on ferromagnetic thin films have measured
the dependence of the hysteresis loop area
on the amplitude and frequency of the external field,
$A$=$A(H_{0},\omega)$,
and approximate agreement with
numerical simulations of Ising models has been reported.
Here we present numerical and theoretical calculations of
$A$ in the low-frequency regime for
two values of $H_{0}$,
which bracket a temperature and system-size
dependent crossover field.
Our previous Monte Carlo
studies have shown that the hysteretic response of the
kinetic Ising model is qualitatively different for
amplitudes above and below this crossover field.
Using droplet theory, we
derive analytic expressions for the low-frequency asymptotic
behavior of the hysteresis loop area.
In both field regimes, the loop area
exhibits an extremely slow
approach to an asymptotic, logarithmic
frequency dependence of the form
$A \propto - [\ln (H_{0} \omega)]^{-1}$.
Our results are relevant to the interpretation of data
from experiments and simulations,
on the basis of which
power-law exponents for the hysteresis-loop area have
been reported.
\end{abstract}
\pacs{PACS number(s):64.60.Qb, 75.60.Ej, 75.10.Hk, 64.60.My}


\narrowtext

When a ferromagnet is subject to an oscillating external field,
$H(t)$=$H_{0} \sin \omega t$,
the time-dependent magnetization, $m(t)$, typically
lags behind the field.
The area of the resulting hysteresis loop,
$A$=$-\oint m(H) \ dH$, equals the energy dissipated per period.
It is therefore frequently measured in studies of periodically
driven magnetic systems.
Recent experiments on
ultrathin ferromagnetic films, \cite{he93,suen97}
as well as numerical simulations of two-dimensional
Ising models, \cite{rao90,lo90,seng92,acha95} have been
interpreted in terms of a low-frequency power law,
$A \propto H_{0}^{a} \omega^{b}$, with
a range of exponent values having been reported. \cite{lo90,seng92,acha95}
This interpretation is not fully consistent with the
fluctuation-free mean-field result, \cite{jung90,luse94}
$A = A_{0} + {\rm const} [\omega^{2} (H_{0}^{2} - H_{\rm sp}^{2})]^{1/3}$
with positive constants
$A_{0}$ and $H_{\rm sp}$, which has been successfully applied to analyze
experiments on ultrathin films of Co on Cu(001). \cite{jiang95}
Nor does the single power-law dependence
agree with the logarithmic dependence expected if
thermally activated nucleation is the
rate-determining process.~\cite{thom93,beal94,kole97}
Here we present analytical and numerical results that indicate a
resolution of this puzzling situation.

Theoretical arguments and numerical simulations
reveal parameter regimes in which,
following instantaneous field reversal,
a uniaxial single-domain ferromagnet
switches to the stable magnetization direction
via two distinct mechanisms.
This magnetization reversal occurs
either by nucleation of a {\it single}
critical droplet of the stable phase [the single-droplet (SD) regime]
or by simultaneous nucleation and growth of {\it many}
critical droplets 
[the multi-droplet (MD) regime].~\cite{kole97,rich_all,rikv94}
The SD~(MD) regime corresponds to weaker~(stronger) fields and/or
smaller~(larger) systems.
In this extension of our previous studies of hysteresis,~\cite{side_all}
we present analytical and Monte Carlo results
for the hysteresis-loop area for a kinetic Ising model
at low frequencies; both in
the SD regime and in the MD regime.
The derivations are based on time-dependent extensions
of classical homogeneous nucleation theory
and ``Avrami's law'' for the
decay of a metastable phase.~\cite{seki86}
For both decay mechanisms we show how
an extremely slow approach of $A$ to an asymptotic logarithmic dependence on
$H_{0} \omega$ as $\omega \rightarrow 0$
gives ``effective exponents'' which superficially
appear to describe a power law,
even for data extending over several decades in frequency.

The model used here is a kinetic, nearest-neighbor
Ising ferromagnet on a
square lattice with
Hamiltonian ${\cal H }$=
$-J \sum_{ {\em \langle ij \rangle}} {\em s_{i}s_{j}} - H(t) \sum_{i} {\em s_{i}}$
and periodic boundary conditions.
Here
$s_{i}$=$\pm 1$ are the local spin variables,
$\sum_{ {\em \langle ij \rangle} }$ runs over all
nearest-neighbor pairs, and $\sum_{i}$ runs over all
$N$=$L^{2}$ lattice sites.
The ferromagnetic exchange coupling is $J>0$, and $H(t)$ is a
time-dependent external field.
The dynamic
is the Glauber single-spin-flip algorithm,
with updates at randomly chosen sites.
It is defined by the spin-flip probability
$W(s_{i} \rightarrow -s_{i})$=
$\exp(- \beta \Delta E_{i})/(1 + \exp(- \beta \Delta E_{i}))$,
where $\Delta E_{i}$ is the change in the energy of the system
if the spin flip is accepted,
and $\beta^{-1} = k_{\rm B}T$ is the temperature in energy units.
Time is given in units of Monte Carlo steps per spin (MCSS).
The average lifetime,
$\langle \tau(|H|) \rangle$, of the
unfavorably magnetized phase in a {\it static} field
of magnitude $|H|$ is defined as the average
time it takes the magnetization to reach zero, following
instantaneous field reversal.
The frequency, $\omega$, of the applied sinusoidal field,
is chosen by specifying the ratio
$R$=$(2 \pi / \omega) / \langle \tau(H_{o}) \rangle$.

We initially prepare a system of size $L$=$64$
at $T$=$0.8T_{c}$ with all spins down, i.e. $m(0)$=$-1$.
Then the sinusoidal field
$H(t)$=$H_{0} \sin \omega t$ is applied,
and $m(t)$ is recorded for a fixed number of MCSS, $n_{\rm max}$.
For the simulations in the SD regime, the field amplitude
is $H_{0}$=$0.1J$ (which gives $\langle \tau \rangle$=$2058$ MCSS)
with $n_{\rm max}$=$16.9 \times 10^{6}$ MCSS.
For the MD regime, the amplitude
is $H_{0}$=$0.3J$ (which gives $\langle \tau \rangle$=$75$ MCSS)
with $n_{\rm max}$=$5.9 \times 10^{5}$ MCSS.
For the values of $L$ and $T$ used here, the
crossover field (called the Dynamic Spinodal \cite{rikv94} (DSP) )
between these two regimes is $H_{\rm DSP} \approx 0.11J$.
For large systems $H_{\rm DSP}$ vanishes slowly with $L$ as
$H_{\rm DSP}(L) \sim (\ln L)^{-1/(d-1)}$.
Figure \ref{fig_loops} shows
representative hysteresis loops from
simulations in both the SD and MD regimes.
The large relative fluctuations in the loop area
in Fig.~\ref{fig_loops}(a) indicate the stochastic
nature of the switching mechanism in the SD regime.~\cite{side98-SDMD}
The relative fluctuations in the loop area are smaller
in the MD regime (Fig.~\ref{fig_loops}(b)).
The stochastic nature of magnetization reversal in the SD regime
allows one to treat the switching as a variable-rate
Poisson process.
For low frequencies, this variable switching rate is the
system volume times the
nucleation rate obtained from classical droplet theory,
 \begin{equation}
 \label{eq_I}
  L^{d} I \Bigl ( H(t),T \Bigr ) 
   \propto 
  L^{d} |H(t)|^K \exp \left [- \frac{\Xi_{0}(T)}{|H(t)|^{d-1}} \right ] ,
 \end{equation}
where $d$ is the spatial dimension of the system, and
$K$ and $\Xi_{0}(T)$
are known from theory and simulations.~\cite{rich_all,rikv94}
The quantity $\Xi_{0}(T)$ is the field-independent part of the free-energy
cost of a critical droplet, divided by $k_{\rm B} T$.
The time dependence of the nucleation rate enters solely through $H(t)$.
Using Eq.~(\ref{eq_I}) one can derive an expression
for the cumulative probability that a switch has taken place
by time $t$, $F(t)$.~\cite{side98-SDMD}
The median switching time, $t_{\rm s}$,
is given by $F(t_{\rm s})$=$1/2$.
To obtain an analytic result we use the low-frequency
approximation $H(t) \approx H_{0} \omega t$.
Then the median switching field,
$H_{\rm s}$=$H_{0} \omega t_{\rm s}$, is given by the solution of the equation,
 \begin{equation}
 \label{eq_gammaSD}
   \ln 2 = 
     \rho_{0}
     \frac{ e^{\frac{\Xi_{0}(T)}{H_{0}^{d-1}}} \Xi_{0}^{\frac{K+1}{d-1}} }
           { H_{0}^{K+1} (d-1) \omega }
     \Gamma \left ( 1-\frac{K+d}{d-1},\frac{\Xi_{0}}{H_{\rm s}^{d-1}} \right ),
 \end{equation}
where $\Gamma(a,x)$ is the incomplete gamma function, and
$\rho_{0}$
is the switching rate in a static field of magnitude $H_{0}$,
which has been measured in field-reversal simulations.~\cite{side98-SDMD}

Due to the square shape of the hysteresis loop in both regimes,
the loop area is given by
$\langle A \rangle / 4 H_{0} \approx m_{\rm eq} H_{\rm s}(\omega)/H_{0}$,
where $m_{\rm eq}$ is the spontaneous zero-field magnetization.
Figure \ref{fig_Afreq_SD} is a log-log plot of the hysteresis-loop
area versus the frequency, $1/R$, in the SD regime.
The solid curve is calculated by numerical solution of
Eq.~(\ref{eq_gammaSD})
with $d$=$2$, $K$=$3$, $\Xi_{0}$=$0.506192J$, and
$\rho_{0}$=$6.62 \times 10^{-4} {\rm MCSS}^{-1}$.
Hence, this calculation involves no adjustable parameters.
The solid dots are data from MC simulations.
Each of the two dashed lines is obtained from a
linear least-squares fit to the numerical solution for the loop area
over nearly four decades in frequency.
The effective exponents obtained from this fitting procedure
appear valid over a frequency range that would be considered large
from the viewpoint of experiments or even simulations.
Over a very large frequency range however, the effective exponent
depends on the frequency range in which
data are analyzed.
Expanding $\Gamma(a,x)$ in Eq.~(\ref{eq_gammaSD}) for large values
of $x$=$\Xi_{0}/H_{\rm s}^{d-1}$ gives the asymptotic low-frequency result
$\langle A \rangle_{\rm SD} \propto - [\ln (H_{0} \omega)]^{-1/(d-1)}$.

The details of the theoretical derivation of the loop area in the
MD regime are different than in the SD regime.
However, two basic features are the same:
the form of the time-dependent nucleation rate,
$I(H(t),T)$ from Eq.~(\ref{eq_I}),
and the linear approximation for the field
used to obtain asymptotic analytic results for very low frequencies.
Figure \ref{fig_Afreq_MD} is a log-log plot of the hysteresis-loop
area versus $1/R$ in the MD regime.
The solid curve results from a full numerical integration (NI)
of an analytic expression for $m(t)$, obtained from
Avrami's law,\cite{seki86} with the sinusoidal form of $H(t)$.
The dotted curve results from a numerical solution (NS)
of an analytic
expression obtained from a linear approximation for $H(t)$,
as for the SD regime.
The transcendental equation that must be solved is analogous to
Eq.~(\ref{eq_gammaSD}),
but contains a sum of three
incomplete gamma functions.~\cite{side98-SDMD}
The MC data (solid dots), NI and NS results are
in excellent agreement.
For $d$=$2$ and {\it extremely} low frequencies, an asymptotic
expansion of the analytic expression used to obtain the
NS result gives
$\langle A \rangle_{\rm MD} \propto - [\ln (H_{0} \omega)]^{-1}$.
As in the SD case, from a log-log plot of the loop area
versus frequency one can extract effective
exponents from the data over nearly two decades in $1/R$.
However, these effective exponents depend strongly on
the frequency range in which the fit is performed.
Similarly, if $A$ is plotted vs. $-[\ln (H_{0} \omega)]^{-1}$
as in Ref.~[10], the slow crossover will result in a
significant overestimate of the asymptotic exponent $1/(d-1)$.

A change from MD to SD
behavior should appear not only
as $H_{0} \rightarrow H_{\rm DSP}$, but
for finite systems it should be observed when $\omega$ becomes
sufficiently low that $H_{\rm s} < H_{\rm DSP}(L)$.
The frequency of this crossover should be given
by the intersection of the results for the loop areas in the SD
and MD regions.
The dashed curve in Fig.~\ref{fig_Afreq_MD} represents
the solid curve in Fig.~\ref{fig_Afreq_SD}, which has been
rescaled so that the two results may be plotted together.
The value of the loop area at the intersection is that
of a loop with $H_{\rm s} \approx H_{\rm DSP}(L)$.
While $m(t)$ and $A$ do not depend on system size in the MD regime,
$A$ in the SD region, and hence the location of the crossover,
depends on $L$.

In conclusion, we have shown that the hysteresis-loop areas
for kinetic Ising ferromagnets driven by oscillating external fields vanish
logarithmically with $H_0 \omega$ for asymptotically low frequencies.
This result should be valid for all fields and temperatures
such that magnetization switching proceeds via a homogeneous
nucleation-and-growth mechanism,~\cite{thom93,beal94}
in particular for both the single-droplet and
multidroplet regimes considered here.
For both of these regimes we stress that
the asymptotic low-frequency behavior would only be seen
for {\it extremely\/} low frequencies.
For frequencies in a more ``realistic'' range we find a wide
crossover, extending over many decades in frequency.
Power-law fits to the loop areas over as much as four frequency decades
give good agreement within the fitting range,
but the resulting effective exponents depend strongly on the fitting interval.
We believe our results are significant to the interpretation and
comparison of results from experimental \cite{he93,suen97,jiang95}
and numerical \cite{lo90,seng92,acha95} studies of
hysteresis in ferromagnetic systems, in which power-law
dependences of the loop areas have been reported
with a variety of exponents.

\acknowledgments{
S.W.S. and P.A.R. thank
P.D. Beale,
G. Brown,
W. Klein,
M. Kolesik, and
R.A. Ramos
for useful discussions
and the Colorado Center for Chaos and Complexity for hospitality
and support during the 1997 Workshop on Nucleation Theory
and Phase Transitions.
Research supported in part by FSU-MARTECH,
by 
FSU-SCRI under DOE
Contract No.\ DE-FC05-85ER25000, 
and by NSF
Grants No.\ DMR-9315969, DMR-9634873, and DMR-9520325.}



\begin{thebibliography}{10}

\bibitem{he93}
Y. He and G. Wang, Phys.\ Rev.\ Lett.\ {\bf 70},  2336  (1993).

\bibitem{suen97}
J.~S. Suen and J. Erskine, Phys.\ Rev.\ Lett. {\bf 78},  3567  (1997).

\bibitem{rao90}
M. Rao, H. Krishnamurthy, and R. Pandit, Phys.\ Rev.\ B {\bf 42},  856  (1990).

\bibitem{lo90}
W. Lo and R.~A. Pelcovits, Phys.\ Rev.\ A {\bf 42},  7471  (1990).

\bibitem{seng92}
S. Sengupta, Y. Marathe, and S. Puri, Phys.\ Rev.\ B {\bf 45},  7828  (1992).

\bibitem{acha95}
M. Acharyya and B.~K. Chakrabarti, Phys.\ Rev.\ B {\bf 52},  6550  (1995), and
  references cited therein.

\bibitem{jung90}
P. Jung, G. Gray, and R. Roy, Phys.\ Rev.\ Lett. {\bf 65},    (1990).

\bibitem{luse94}
C. Luse and A. Zangwill, Phys.\ Rev.\ E {\bf 50},  224  (1994).

\bibitem{jiang95}
Q. Jiang, H.-N. Yang, and G.-C. Wang, Phys.\ Rev.\ B {\bf 52},  14911  (1995).

\bibitem{thom93}
P.~B. Thomas and D. Dhar, J.\ Phys.\ A: Math.\ Gen.\ {\bf 26},  3973  (1993).

\bibitem{beal94}
P.~D. Beale, Integrated Ferroelectrics {\bf 4},  107  (1994).

\bibitem{kole97}
M. Kolesik, M.~A. Novotny, and P.~A. Rikvold, Phys.\ Rev.\ B {\bf 56},  11791
  (1997).

\bibitem{rich_all}
H.~L. Richards, S.~W. Sides, M.~A. Novotny, and P.~A. Rikvold, 
 J.\ Magn.\ Magn.\ Mater.\ {\bf 150},  37  (1995);
 Phys.\ Rev.\ B {\bf 54}, 4113  (1996);
 Phys.\ Rev.\ B {\bf 55},  11521  (1997).

\bibitem{rikv94}
P.~A. Rikvold, H. Tomita, S. Miyashita, and S.~W. Sides, Phys.\ Rev.\ E {\bf
  49},  5080  (1994).

\bibitem{side_all}
S.~W. Sides, R.~A. Ramos, P.~A. Rikvold, and M.~A. Novotny,
 J.\ Appl.\ Phys.\ {\bf 79},  6482  (1996);
 J.\ Appl.\ Phys.\ {\bf 81},  5597  (1997).

\bibitem{seki86}
K. Sekimoto, Physica {\bf 135A},  328  (1986).

\bibitem{side98-SDMD}
S.~W. Sides, P.~A. Rikvold, and M.~A. Novotny, In preparation  (1998).

\end{thebibliography}


\begin{figure}
\caption{\label{fig_loops}
Low-frequency hysteresis loops from simulations of
a kinetic Ising model.
For both regimes five loops are shown, representing
short portions of the entire simulation time series.
(a) Loops from the single-droplet (SD) regime,
using $H_{0}$=$0.1J$ at a scaled frequency of $1/R$=$0.01$.
(b) Loops from the multi-droplet (MD) regime, 
using $H_{0}$=$0.3J$ at a scaled frequency of $1/R$=$0.005$.
}
\end{figure}

\begin{figure}
\caption{\label{fig_Afreq_SD}
Log-log plot of $\langle A \rangle / 4H_{0}$ vs. $1/R$
in the SD regime.
The solid curve is obtained from
the numerical solution of Eq.~(\ref{eq_gammaSD}),
the derivation of which is outlined in the text.
The dashed line segments represent linear least-squares fits
to different portions of the numerical solution data.
The data that yield the effective exponent
$b$=$0.096$, are centered around $\log (1/R)$=$-2.05$;
those that yield $b$=$0.033$ are centered around $\log (1/R)$=$-13.38$.
The two solid dots are MC simulation data.
The vertical lines are not error bars; they
represent the standard deviation of the loop-area distribution.
}
\end{figure}

\begin{figure}
\caption{\label{fig_Afreq_MD}
Log-log plot of $\langle A \rangle /4H_{0}$ vs. $1/R$ in
the MD regime.
The solid curve is obtained from a full numerical integration of the
time-dependent Avrami's law result for $m(t)$, using a sinusoidal field,
$H(t)$=$H_{0} \sin \omega t$.
(This calculation could not be extended to
lower frequencies than those shown due to numerical difficulties.)
The dotted curve is obtained from a numerical solution of
an analytic expression whose derivation uses
a linear approximation for the field, $H(t) \approx H_{0} \omega t$.
The solid dots represent MC simulations.
The vertical bars denote the standard deviation in the loop-area
distributions as in Fig.~\protect\ref{fig_Afreq_SD}.
The dashed curve represents the SD result (the solid curve
in Fig.~\protect\ref{fig_Afreq_SD}) after rescaling so that
the SD and MD results may be compared.
}
\end{figure}
\pagebreak


\begin{figure}
\noindent
Figure 1
\hspace{5.0in}
{\large {\bf CD-06}} \\
S.W. Sides \\
MMM-Intermag 98 \\
\vspace{0.75in}
\centerline{\psfig{figure=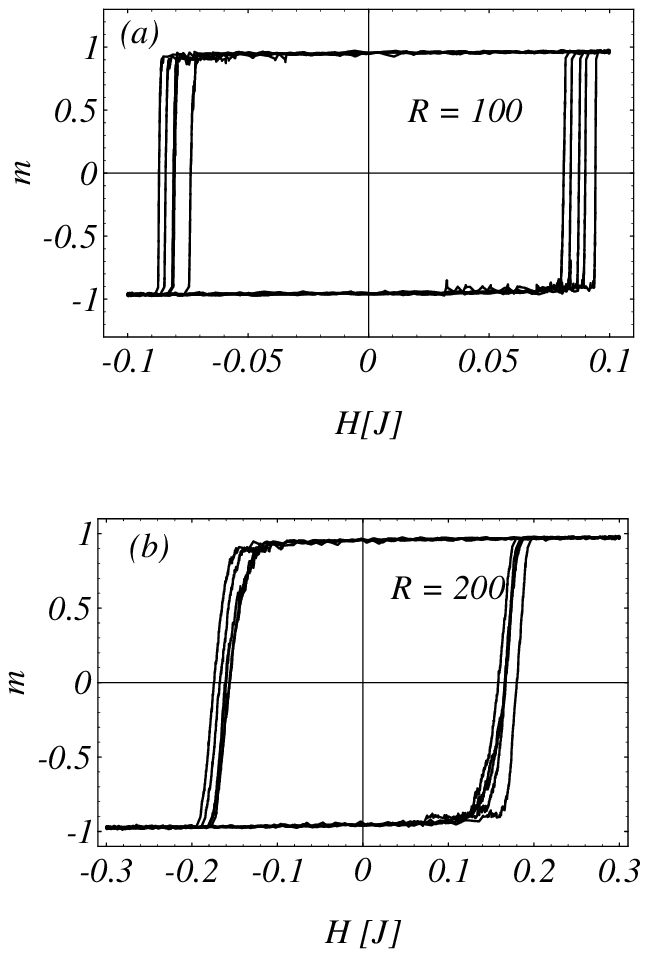,width=5.5in}}
\end{figure}
\pagebreak

\begin{figure}
\noindent
Figure 2
\hspace{5.0in}
{\large {\bf CD-06}} \\
S.W. Sides \\
MMM-Intermag 98 \\
\vspace{0.75in}
\centerline{\psfig{figure=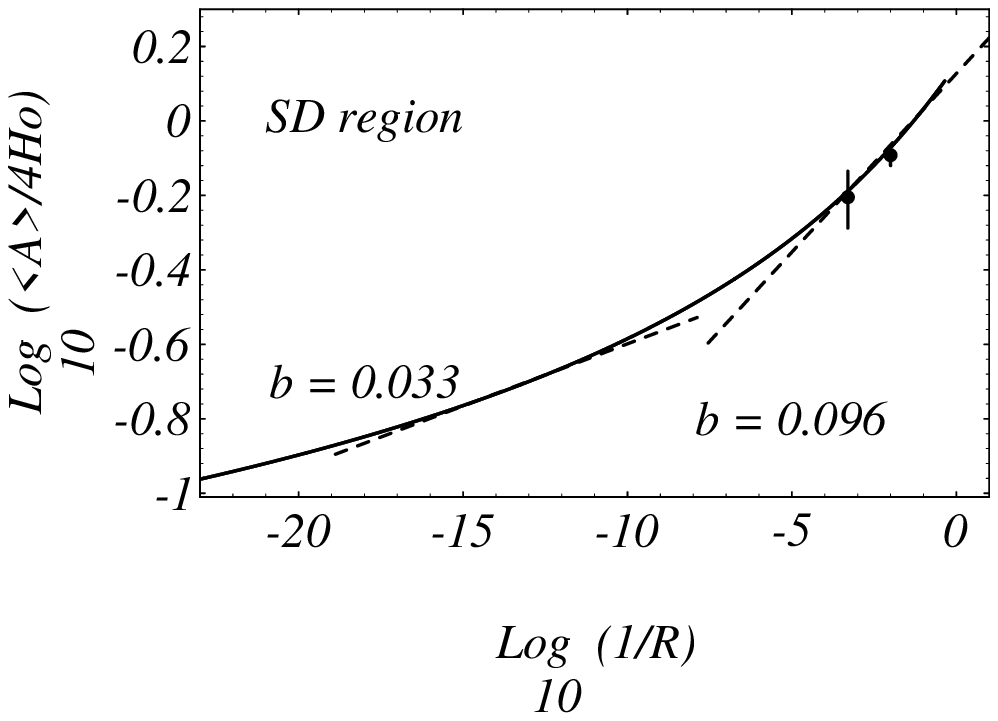,width=3.6in}}
\end{figure}
\pagebreak

\begin{figure}
\noindent
Figure 3
\hspace{5.0in}
{\large {\bf CD-06}} \\
S.W. Sides \\
MMM-Intermag 98 \\
\vspace{0.75in}
\centerline{\psfig{figure=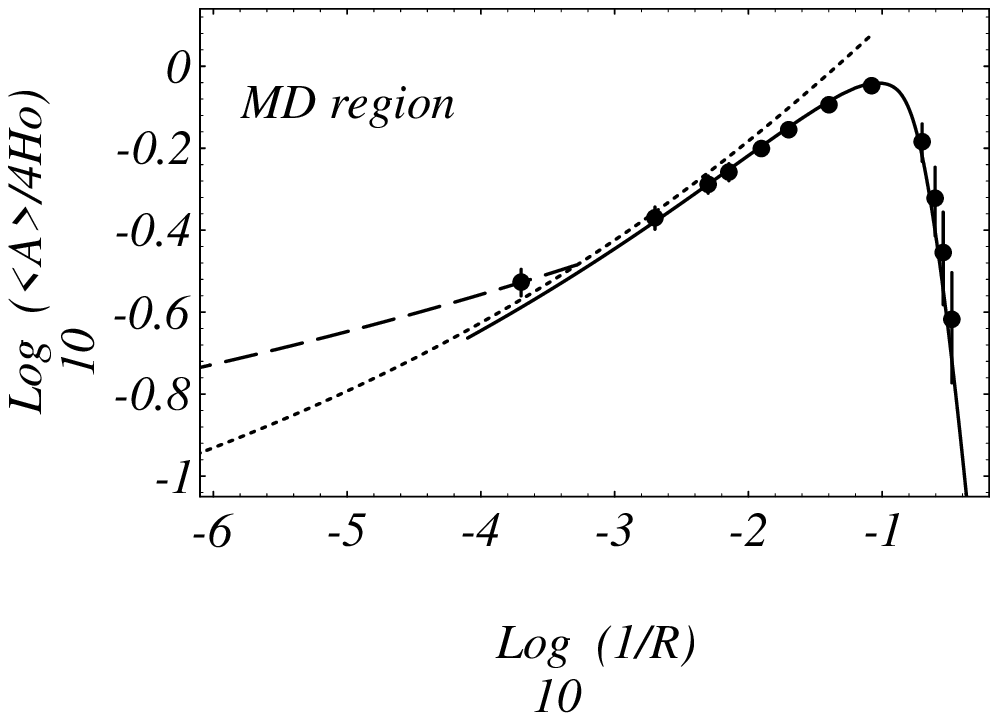,width=3.6in}}
\end{figure}
\pagebreak


\end{document}